# Quasi-Fractal UCA Based OAM for Highly Efficient Orthogonal Transmission

Wenchi Cheng, *Senior Member, IEEE*, Haiyue Jing, *Student Member, IEEE*, Wei Zhang, *Fellow, IEEE*, Keyi Zhang, *Student Member, IEEE*, and Hailin Zhang, *Member, IEEE*

*Abstract*— The development of orbital angular momentum (OAM)-based radio vortex transmission presents a promising opportunity for increasing the capacity of wireless communication in correlated channels due to its inherent orthogonality among different OAM modes. One of the most popular schemes for high-efficient OAM transmission is the digital baseband associated with uniform circular array (UCA) based transceiver. However, the periodicity of complex-exponential feed makes the maximum number of orthogonal signals carried by multiple OAM modes generally restricted to the array-element number of UCA antenna, which poses an open question of how to employ more OAM modes given a fixed number of array elements. Furthermore, signals modulated with high-order OAM modes are difficult to be captured by the receiver due to their serious divergence as propagating in free space, thus severely limiting the capacity of radio vortex communications. To overcome the above challenges, in this paper based on the partly element-overlapped fractal geometry layout and effectively using low-order OAM modes, we propose the quasi-fractal UCA (QF-UCA) antenna based OAM multiplexing transmission. We perform the two-dimension OAM modulation (TOM) and demodulation (TOD) schemes with the orthogonal OAM mode number exceeding the array-element number, which is beyond the traditional concept of multiple antennas based wireless communications. Simulation results show that our proposed scheme can achieve more number of orthogonal multiplexing streams than the maximum number of orthogonal multiplexing corresponding to traditional multiple antenna systems.

*Index Terms*— Orbital angular momentum (OAM), quasi-fractal UCA (QF-UCA), two-dimension OAM modulation (TOM), mode-number over element-number.

## I. Introduction

RESEARCHERS engaged in the field of wireless communications consistently focus on achieving optimal spectrum efficiency (SE). The rapid emergence of new communication services and applications has led to an unprecedented growth in data traffic. However, this has also resulted in a shortage of radio spectrum resources, creating a significant challenge for the development of high-capacity data transmission in the future [1], [2]. In recent decades, a number of novel transmission technologies have sprung up continuously, providing effective and reliable schemes to further enhance the capacity of wireless communications, such as millimeter wave (mmWave) [3], terahertz wave [4], ultra-dense networking [5], massive multiple-input multiple-output (MIMO) [6], co-frequency co-time full-duplex [7], reconfigurable intelligent surfaces [8], etc. since high frequency causes severe attenuations [9]. The orbital angular momentum (OAM) radio vortex transmission over line-of-sight (LOS) channels have received much attentions [10]–[12]. Moreover, the LOS transmission is critical as it encounters substantial attenuation due to the high frequency [9]. Consequently, there is significant interest in the use of OAM radio vortex transmission for LOS channels [10]–[12].

The fundamental physical quantity of radio vortex waves, referred to as OAM, was initially discovered in field of optics, but has applied to various fields, including wireless telecommunications [13], radar detection [14], and other microwave-based technologies [15]. The order of OAM modes theoretically takes an infinite number of values [16]–[18]. Theoretically, the sum capacity and SE of OAM-based radio vortex are proportional to the number of OAM modes transmitted simultaneously at the same frequency, which offers a brand-new perspective on multiplexing transmission, while also facilitating seamless cooperation with existing techniques to effectively enhance the overall system performance of wireless communications [19]–[23]. In addition, since the integration of antennas operating at the mmWave frequency band into large arrays for multi-data transmission creates potential for significantly improving SE [24], the mmWave radio vortex technology is expected to greatly elevate the level of achievable SE in future wireless communications [25], [26].

Various experimental methods have been demonstrated for the generation of radio vortex signals (OAM signals) in terms of signal transmission. Spiral phase plate (SPP) was proposed and validated in optics and mmWave domain [27]. The parabolic antenna is presented at relatively low frequencies [28] and the electromagnetic metasurface is designed to generate OAM signals [29]. In addition, the uniform circular array (UCA) antenna has been widely used to generate and receive the OAM signals [30]–[36]. As a typical kind of array antenna, one UCA is equipped with multiple array-elements distributed around the substrate circumference. In contrast to other antenna structures, UCA antennas offer the advantages of enhanced flexibility, reduced space requirements, and simplified adjustment. There have been extensive research works focusing on investigating the UCA based radio vortex communications [20], [37]–[42]. The authors of [37] analyzed the channel capacity of radio vortex transmission with UCA antennas and the OAM modulation combined with orthogonal

W. Cheng, H. Jing, K. Zhang, and H. Zhang are with the State Key Laboratory of Integrated Services Networks, Xidian University, Xi'an, 710071, China (e-mails: wccheng@xidian.edu.cn; hyjing@stu.xidian.edu.cn; kyzhang@stu.xidian.edu.cn; hlzhang@xidian.edu.cn).

W. Zhang is with School of Electrical Engineering and Telecommunications, the University of New South Wales, Sydney, Australia (wzhang@ee.unsw.edu.au ).

This work was supported in part by the National Natural Science Foundation of China (No.62341132) and in part by QTZX24078.

frequency-division multiplexing (OFDM) is introduced in [20], [38]. For extensive environments, the authors of [39] have studied the multipath impact on OAM signals and the authors of [40] have characterized the high-order statistics of OAM modes under atmosphere turbulence.

For UCA based radio vortex transmission, multiple coaxial data streams can be transmitted, where each OAM mode carries one independent data stream (called OAM multiplexing) [16]. Typically, equal amplitude and specific phase differences are feed on the UCA antenna to generate different OAM modes carrying emitted signals. According to the periodicity of complex exponential, the maximum number of signals with different OAM modes is restricted to the array-element number of UCA antenna, which leads to an upper limit of the capacity [34], [37], [43]. Limited research works have been conducted regarding the effective multiplexing of a larger number of OAM modes beyond the allocated array-elements associated with the UCA, challenging the conventional concept of the multiple antennas based wireless communications. In addition, with an increasing order of OAM mode, the beam experiences more intense divergence, resulting in a sharp decrease in the energy captured at the receiver. Hence, it is challenging to modulate the transmit signals on high-order OAM modes [44]. Therefore, how to explore the way of increasing the number of available OAM modes is a crucial and unresolved problem.

To address the above-mentioned problems, in this paper we propose the quasi-fractal UCA (QF-UCA) antenna based two-dimension OAM multiplexing to generate more orthogonal OAM modes, i.e., the number of OAM modes is larger than that of array-elements. We design a new geometry layout of antenna and use the geometric axisymmetry of circular arrays together with the fractal geometry principle to implement OAM multiplexing in a new way. First, we design a single-loop UCA based two-dimension QF antenna layout, where each UCA is regarded as an integral unit (hereinafter called the UCA cell) and the array-elements at the overlapping azimuth synchronously play different roles in generating multiple OAM signals with partial fractal deployment. Then, based on the QF antenna, we develop Two-dimension Discrete Fourier Transformation (DFT) based OAM Modulation (TOM) and Demodulation (TOD) schemes for multiple OAM modes transmissions, where two independent sets of low-order OAM modes are selected for multiplexing within two different coordinate systems. Also presented are numerical simulations of our system performance under 5.8GHz since it is a popular frequency for wireless communications, showing that our designed QF antenna layout and proposed TOM/TOD schemes can significantly increase the spectrum efficiency as compared with traditional multiple antennas based wireless communications. Generally, our proposed QF-UCA antenna structure and the corresponding schemes can be adapted at other frequencies.

The remainder of this paper is organized as follows. The system model and the geometric model of QF-UCA antenna are given in Section II. TOM and TOD schemes are analyzed in Section III, along with the equivalent channel model. Performance evaluations are shown in Section IV and the paper concludes with Section V.

*Notation*: Matrices and vectors are denoted by the capital letters and the lowercase letters in bold, respectively. We generalize the definition of traditional matrix with the notation from $\mathbb{C}^{M \times N}$ to $\mathbb{C}^{(M \times N)\{V \times K\}}$, which represents that elements of one matrix can be matrices, vectors ($K=1$), and numerical values ($V = K = 1$). The notation $\times$ represents the matrix multiplication, $\cdot$ represents the element-wise matrix multiplication where the element is generalized (matrix, vector, or one-dimension number), $((\cdot))_N$ denotes the modulus operation with divisor $N$, and $\text{diag}(\boldsymbol{\alpha})$ represents a square diagonal matrix with the elements of vector $\boldsymbol{x}$ on the main diagonal. The notation $\text{diag}(\boldsymbol{X})$ represents a vector of main diagonal elements of matrix $\boldsymbol{X}$. The notations $(\cdot)^H$, $(\cdot)^{-1}$, $(\cdot)^T$, and $(\cdot)^*$ denote the Hermitian, the inverse, the transpose, and the conjugation of a matrix or a vector, respectively.

## II. SYSTEM MODEL

Figure 1 depicts the system model for OAM multiplexing transmission with QF-UCA antennas on both transmitter and receiver. The transmitter contains the inner-UCA modulation, inter-UCA modulation, power combiner, and transmitted QF-UCA antenna. The inner-UCA modulation and the inter-UCA modulation forms the two-dimension inverse Discrete Fourier Transformation (IDFT) based OAM modulation for input signals. The power combiner along with transmitted QF-UCA antenna are used for integrating and emitting multiple OAM modes signals. The receiver contains received QF-UCA antenna, power splitter, inter-UCA demodulation, post-decoding, inner-UCA demodulation, and $M$-branch mode-wise detection. The received QF-UCA antenna and power splitter are used for receiving and separating OAM modes signals. The inter-UCA demodulation, the inner-UCA demodulation, and the post-decoding are used to recover the orthogonality of OAM modes.

Here, we propose to use the UCA-based quasi-fractal antenna for both transmitter and receiver. For the transmitted QF-UCA antenna, there are $N_t$ patches which are divided into $N$ UCA cells and $K$ patches are equipped in each UCA cell. Each UCA cell can transmit $K$ data streams and the number of data streams is $NK$ when there are $N$ UCA cells. Since there exist reused patches among different UCA cells, the total number of transmitted data streams is larger than the number of patches equipped in the QF-UCA, i.e., $NK > N_t$. The received antenna is also grouped, where we denote by $N_r$, $M$, and $V$ the numbers of all received patches, UCA cells, and patches per UCA cell, respectively. UCA cells indexed with $n$ at the transmitter and UCA cells indexed with $m$ at the receiver are uniformly distributed along the circumference of entire substrate. The inner patches indexed with $k$ and $v$ at the transmitter and receiver, respectively, are also equidistantly distributed along the circumference of each UCA cell. Notations $R_t$, $R_r$, and $R_Q$ represent the radii of transmit UCA cell, receive UCA cell, and QF-UCA antennas, respectively. In addition, the value of $R_Q$ is required to be the same for the transmitter and the receiver. In the inner-UCA coordinate system, x-axis is set as the direction from the center of UCA cell to the first patch ($k, v = 0$) while z-axis is the centering normal line pointing to the directly opposite receive UCA cell. In the inter-UCA coordinate system, x-axis is set as the direction from the center of substrate to the first UCA cell named $UCA^0$ while z-axis is the centering normal line pointing to the received QF-UCA antenna. Correspondingly, y-axis are decided by the right-hand spiral rule. Based on this, the position of each patch in three-dimension (3D) space can be described by the two coordinates, where $\tilde{\theta}_k$ and $\theta_v$ denote the azimuth within the UCA cell at the transmitter and receiver, respectively. The notation $\tilde{\Theta}_n$ and $\Theta_m$ denote the azimuth of the UCA cells involved at the transmitter and receiver, respectively. In this paper, we assume that the transmitted and received QF-UCA antennas are strictly aligned with each other and the design-related requirements for antenna, such as the impedance matching and the spatial correlation, are well satisfied. Also, we mainly consider LOS channel in this paper.

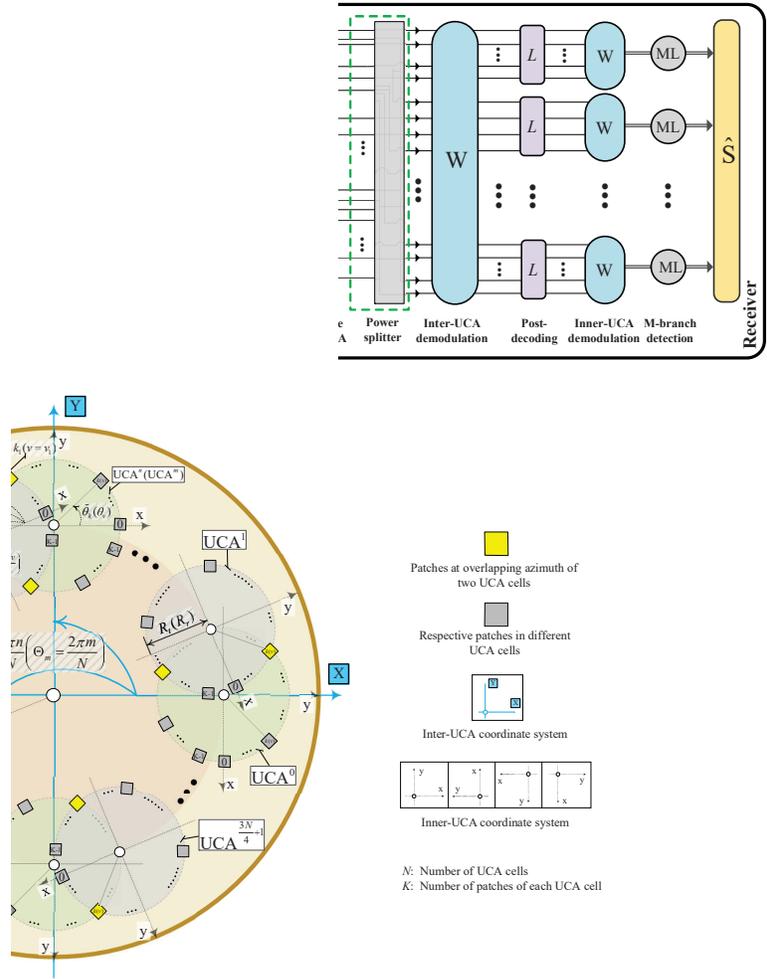

Fig. 1. The system model for OAM multiplexing transmission with the UCA based quasi-fractal array antennas.

As illustrated in Fig. 1, signals in the same branch are first given a specified phase gradient $2\pi l/K$, where $l$ ($1-\lfloor K/2 \rfloor \leq l \leq \lfloor K/2 \rfloor$) denotes the order of OAM mode used for the inner-UCA (the first-dimension) modulation, $\lfloor \cdot \rfloor$ represents the round down operation, and the constant factor $2\pi/K$ matches the azimuth difference of any two adjacent array-elements within the UCA cell. This process is equivalent to applying the $K$-point IDFT for the $K$ input signals. The inter-UCA (the second-dimension) modulation is performed that the signals corresponding to the same UCA cell are with a phase gradient $2\pi p/N$, where $p$ ($1-\lfloor N/2 \rfloor \leq p \leq \lfloor N/2 \rfloor$) denotes the order of OAM mode used for the inter-UCA modulation and the constant $2\pi/N$ matches the azimuth difference of any two adjacent UCA cells. This process is equal to the $N$-point IDFT for $N$ signal vectors. Thus, there is a modulation factor $e^{j\frac{2\pi lk}{K}} e^{j\frac{2\pi pn}{N}} / \sqrt{KN}$ for one input source to the patch indexed with $k$ corresponding to the transmit UCA cell indexed with $n$. For the signals transmitted by the $UCA^n$ cell, there are $M$ subchannels $\boldsymbol{H}_{m,n} \in \mathbb{C}^{(V \times K)}$, each of which contains $V$ paths. At the receiver, received signals are first split into $M$ branches. The first-dimension demodulation is to separate the inter-UCA OAM modes carrying signals with performing phase compensation based on the azimuths of receive UCA cells, which is the process of $M$-point IDFT for $V$ signal vectors. Regarding the inner-UCA OAM modes within each UCA cell, the channel model can be considered as the combination of channels from the transmit cells with index $n$ and the receive UCA cells with index $m$, which are with coaxial and non-coaxial [45], after the first-dimension demodulation. We derive the channel models corresponding to different OAM modes and cancel the interference among different OAM modes, i.e., the signals regarding each branch can be split into $V$ streams. Then, the signals carried by different OAM modes are demodulated without interference according to the orthogonality among OAM modes.

## III. Two-dimension OAM Multiplexing for QF-UCA Antenna Based Radio Vortex Transmission

In this section, the two-dimension OAM multiplexing including TOM scheme and TOD scheme is developed to achieve orthogonal OAM modes. We assume that the number of OAM modes is larger than the number of array-elements and we focus on the LOS scenario while the transmitted and received QF-UCA antennas are aligned with each other. The scenario where the transmitter and receiver are misaligned for the QF-UCA based OAM multiplexing will be investigated in the future.

### A. Two-dimension IDFT Based OAM Modulation (TOM)

We denote by $s_{p,l}$ the transmit signal corresponding to inner-UCA OAM mode $l$ and inter-UCA OAM mode $p$. Then, for the non-shared array-elements, the modulated signal on the array-element indexed with $k$ of transmit UCA$^n$ cell, represented by $x_{n,k}$, can be given as follows:

$$x_{n,k} = \sum_{p=1-\lfloor N/2 \rfloor}^{\lfloor N/2 \rfloor} \sum_{l=1-\lfloor K/2 \rfloor}^{\lfloor K/2 \rfloor} \frac{s_{p,l}}{\sqrt{NK}} e^{j\frac{2\pi k}{K}l} e^{j\frac{2\pi n}{N}p}, \quad (1)$$

where $0 \leq k \leq K-1$, $0 \leq n \leq N-1$, $e^{j\frac{2\pi kl}{K}}$ along with $e^{j\frac{2\pi np}{N}}$ are the phase shifts of inner-UCA and inter-UCA modulations, respectively, and $NK$ is the number of available OAM modes. Since the periods of the items $e^{j\frac{2\pi k}{K}l}$ and $e^{j\frac{2\pi n}{N}p}$ are $2\pi$, respectively, the sets $\{e^{j\frac{2\pi k}{K}l}e^{j\frac{2\pi n}{N}p}| 1-\lfloor K/2 \rfloor \leq k \leq \lfloor K/2 \rfloor, 1-\lfloor N/2 \rfloor \leq n \leq \lfloor N/2 \rfloor\}$, $\{e^{j\frac{2\pi k}{K}l}e^{j\frac{2\pi n}{N}p}| 0 \leq k \leq K-1, 0 \leq n \leq N-1\}$, and $\{e^{j\frac{2\pi k}{K}l}e^{j\frac{2\pi n}{N}p}|K \leq k \leq 2K-1, N \leq n \leq 2N-1\}$ are the same. It is clear that $NK$ is larger than $N_t$, which is the number of array-elements equipped for the transmitted QF-UCA. When there are two shared patches between two adjacent UCA cells and there is no shared patch between two non-adjacent UCA cells, we can obtain that $N_t = NK - 2N$. For the shared array-elements indexed with different $k$ of different transmit UCA, the modulated signal can be written as follows:

$$\underbrace{x_{n_1,k_1} = x_{n_2,k_2} = \cdots = x_{n_U,k_U}}_{U} =$$
$$\sum_{u=1}^{U} \sum_{p=1-\lfloor N/2 \rfloor}^{\lfloor N/2 \rfloor} \sum_{l=1-\lfloor K/2 \rfloor}^{\lfloor K/2 \rfloor} \frac{s_{p,l}}{\sqrt{NK}} e^{j\frac{2\pi k_u}{K}l} e^{j\frac{2\pi n_u}{N}p}, (2)$$

where $U$ is the sharing frequency of the shared array-element and $x_{n_u,k_u}$ ($1 \leq u \leq U$) represents the superposition of the modulated signals indexed with different $k$ and $n$ for the same shared array-element.

Then, we denote by $\boldsymbol{x}_n$ the signal vector for UCA$^n$ cell and we have $NK$ signals, denoted by $\boldsymbol{X} = [\boldsymbol{x}_0 \cdots \boldsymbol{x}_n \cdots \boldsymbol{x}_{N-1}]^T$, for the transmitted QF-UCA antenna as shown in (3). In (3), $\boldsymbol{s}_p = [s_{p,0} \cdots s_{p,l} \cdots s_{p,K-1}]^T$, $\boldsymbol{W} \in \mathbb{C}^{(K \times K)}$ is the $K$-point IDFT matrix, $\boldsymbol{W}_N \in \mathbb{C}^{(N \times N)}$ denotes the $N$-point IDFT matrix, the matrix $\dot{\boldsymbol{W}}_N \in \mathbb{C}^{(N \times N)\{K \times K\}}$ corresponds to the two-dimension IDFT based OAM modulation for signals $\boldsymbol{S} = [\boldsymbol{s}_0 \cdots \boldsymbol{s}_p \cdots \boldsymbol{s}_{N-1}]^T$, and the matrix $\boldsymbol{T}_t$ represents the superpose matrix corresponding to the shared array-elements at the transmitter. The aforementioned operations are hereby referred to as TOM for OAM multiplexing transmissions. Using our proposed TOM scheme, QF-UCA antenna can generate multiple OAM modes and the number of OAM modes is larger than that of array-elements.

### B. Wireless Channel for QF-UCA Based OAM Transmission

Generally, we assume that the number of UCA cells for transmitter and receiver satisfies $M = N$ while $K$ and $V$ can be different. We denote by $h_{v,k}^{m,n}$ the complex channel gain from the array-element indexed with $k$ on transmit UCA$^n$ cell to the array-element indexed with $v$ on receive UCA$^m$ cell. Based on the path loss of radio waves in free space [46], $h_{v,k}^{m,n}$ can be written as follows:

$$h_{v,k}^{m,n} = \frac{1}{L_v} \cdot \frac{\beta\lambda}{4\pi} \cdot \frac{e^{-j\frac{2\pi}{\lambda}d_{v,k}^{m,n}}}{d_{v,k}^{m,n}}, \quad (4)$$

where $\beta$ denotes the parameter gathering relevant constants on antenna elements and their patterns, $\lambda$ represents the radio wavelength, $d_{v,k}^{m,n}$ is the transmission distance from the transmit array-element indexed with $k$ of UCA$^n$ to the received array-element indexed with $v$ of UCA$^m$, and $L_v$ denotes the sharing frequency of received array-element indexed with $v$. For each UCA cell, we have the same sharing frequency matrix, the value of which is given by the following Lemma 1.

*Lemma 1:* The sharing frequency matrix $\boldsymbol{L}$ for each UCA cell is determined by the value of $N$ and $R/R_Q$ (here we denote by $R$ the radius of UCA cell). The lower bound of $\boldsymbol{L}$ is an identity matrix $\boldsymbol{I}$.

*Proof:* The proof is provided in Appendix A. ∎

We denote by $\boldsymbol{H}^F$ and $\boldsymbol{H}_{m,n}$ the channel matrix from the transmitter to the receiver and the submatrix corresponding to the sub-channel from transmit UCA$^n$ cell to receive UCA$^m$ cell, respectively. Thus, the wireless channel of OAM

$$\boldsymbol{X} = \begin{bmatrix} \boldsymbol{x}_0 \\ \vdots \\ \boldsymbol{x}_n \\ \vdots \\ \boldsymbol{x}_{N-1} \end{bmatrix} = \frac{\boldsymbol{T}_t}{\sqrt{N}} \underbrace{\begin{bmatrix} \boldsymbol{E} & \cdots & \boldsymbol{E} & \cdots & \boldsymbol{E} \\ \vdots & \ddots & \vdots & \ddots & \vdots \\ \boldsymbol{E} & \cdots & e^{j\frac{2\pi np}{N}}\boldsymbol{E} & \cdots & e^{j\frac{2\pi n(N-1)}{N}}\boldsymbol{E} \\ \vdots & \ddots & \vdots & \ddots & \vdots \\ \boldsymbol{E} & \cdots & e^{j\frac{2\pi(N-1)p}{N}}\boldsymbol{E} & \cdots & e^{j\frac{2\pi(N-1)^2}{N}}\boldsymbol{E} \end{bmatrix}}_{\boldsymbol{W}_N} \begin{bmatrix} \boldsymbol{W}\boldsymbol{s}_0 \\ \vdots \\ \boldsymbol{W}\boldsymbol{s}_p \\ \vdots \\ \boldsymbol{W}\boldsymbol{s}_{N-1} \end{bmatrix} = \frac{\boldsymbol{T}_t}{\sqrt{N}} \underbrace{\begin{bmatrix} \boldsymbol{W} & \boldsymbol{W} & \cdots & \boldsymbol{W} \\ \vdots & \vdots & \ddots & \vdots \\ \boldsymbol{W} & e^{j\frac{2\pi n}{N}}\boldsymbol{W} & \cdots & e^{j\frac{2\pi n(N-1)}{N}}\boldsymbol{W} \\ \vdots & \vdots & \ddots & \vdots \\ \boldsymbol{W} & e^{j\frac{2\pi(N-1)}{N}}\boldsymbol{W} & \cdots & e^{j\frac{2\pi(N-1)^2}{N}}\boldsymbol{W} \end{bmatrix}}_{\dot{\boldsymbol{W}}_N} \begin{bmatrix} \boldsymbol{s}_0 \\ \vdots \\ \boldsymbol{s}_p \\ \vdots \\ \boldsymbol{s}_{N-1} \end{bmatrix}. \quad (3)$$

$$\begin{aligned}
d_{q,v,k} &= \sqrt{D^2 + [R_r \cos\phi_v - R_t \cos(\psi_k+\varphi_q) - a_q]^2 + [R_r \sin\phi_v - R_t \sin(\psi_k+\varphi_q) - b_q]^2} \\
&= \Big\{ D^2 + [R_r \cos\phi_v - R_t \cos(\psi_k + \varphi_q) + R_Q \sin\varphi_q]^2 \\
&\qquad\qquad\qquad\qquad + [R_r \sin\phi_v - R_t \sin(\psi_k + \varphi_q) + R_Q(1 - \cos\varphi_q)]^2 \Big\}^{\frac{1}{2}} \\
&= \Big[ D^2 + 2R_Q^2 + R_r^2 + R_t^2 + 4R_Q R_r \sin\tfrac{\varphi_q}{2}\cos(\phi_v - \tfrac{\varphi_q}{2}) - 4R_Q R_t \sin\tfrac{\varphi_q}{2}\cos(\psi_k + \tfrac{\varphi_q}{2}) \\
&\qquad\qquad\qquad\qquad - 2R_r R_t \cos(\psi_k + \varphi_q - \phi_v) - 2R_Q^2 \cos\varphi_q \Big]^{\frac{1}{2}} \\
&\approx D + \frac{2R_Q^2 + R_r^2 + R_t^2}{2D} + \frac{2R_Q R_r \sin\tfrac{\varphi_q}{2}\cos(\phi_v - \tfrac{\varphi_q}{2}) - R_Q^2 \cos\varphi_q}{D} \\
&\qquad\qquad - \frac{2R_Q R_t \sin\tfrac{\varphi_q}{2}\cos(\psi_k + \tfrac{\varphi_q}{2}) + R_r R_t \cos(\psi_k + \varphi_q - \phi_v)}{D} \\
&= D + \frac{2R_Q^2 + R_r^2 + R_t^2}{2D} + \frac{2R_Q R_r \sin\tfrac{\varphi_q}{2}\cos(\phi_v - \tfrac{\varphi_q}{2}) - R_Q^2 \cos\varphi_q}{D} \\
&\qquad - \frac{R_t \sqrt{4R_Q^2 \sin^2\tfrac{\varphi_q}{2} + 4R_Q R_r \sin\tfrac{\varphi_q}{2}\cos(\phi_v - \tfrac{\varphi_q}{2}) + R_r^2}}{D} \cos(\psi_k - \phi_v + \varphi_q + \alpha_{q,v}). \quad (10)
\end{aligned}$$

---

based transmission with QF-UCA antennas, denoted by $\boldsymbol{H} \in \mathbb{C}^{(N \times N)\{V \times K\}}$, can be derived as follows:

$$\boldsymbol{H} = \boldsymbol{T}_r \boldsymbol{H}^F \boldsymbol{T}_t = \begin{bmatrix} \boldsymbol{H}_{0,0} & \cdots & \boldsymbol{H}_{0,n} & \cdots & \boldsymbol{H}_{0,N-1} \\ \vdots & \ddots & \vdots & \ddots & \vdots \\ \boldsymbol{H}_{m,0} & \cdots & \boldsymbol{H}_{m,n} & \cdots & \boldsymbol{H}_{m,N-1} \\ \vdots & \ddots & \vdots & \ddots & \vdots \\ \boldsymbol{H}_{M-1,0} & \cdots & \boldsymbol{H}_{M-1,n} & \cdots & \boldsymbol{H}_{M-1,N-1} \end{bmatrix}, \quad (5)$$

where $\boldsymbol{T}_r$ is the superpose matrix corresponding to the shared array-elements at the receiver. Due to the circulant property in $\boldsymbol{H}$, we use a subscript $q$, which is derived from the UCA-related indices $m$ and $n$ that $q = ((n+N-m))_N$. We denote by $\boldsymbol{H}_q \in \mathbb{C}^{(V \times K)}$ the submatrix corresponding to the sub-channel between a pair of transmit and receive UCA cells with indices satisfying $q = ((n+N-m))_N$. Then, the circulant partitioned matrix $\boldsymbol{H} = Circ(\boldsymbol{H}_q)(q = 0, 1, \cdots, N-1)$ can be rewritten as follows:

$$\boldsymbol{H} = \begin{bmatrix} \boldsymbol{H}_0 & \cdots & \boldsymbol{H}_q & \cdots & \boldsymbol{H}_{N-1} \\ \boldsymbol{H}_{N-1} & \boldsymbol{H}_0 & \cdots & & \boldsymbol{H}_{N-2} \\ \vdots & \vdots & \ddots & & \vdots \\ \boldsymbol{H}_1 & \boldsymbol{H}_2 & \cdots & & \boldsymbol{H}_0 \end{bmatrix}. \quad (6)$$

Figure 2 shows the equivalent channel gains for transmit signals with 9-element QF-UCA array antennas. It is obvious that the circulant partitioned matrix $\boldsymbol{H}$ is tractable so that we can focus on the change of signals emitted from a certain UCA cell separately.

Whereafter, we discuss the sub-channel matrix $\boldsymbol{H}_q$ containing $VK$ transmission paths. We rewrite (4) as follows:

$$h_{q,v,k} = \frac{1}{L_v} \cdot \frac{\beta\lambda}{4\pi} \cdot \frac{e^{-j\frac{2\pi}{\lambda}d_{q,v,k}}}{d_{q,v,k}}, \quad (7)$$

where $d_{q,v,k} = d_{v,k}^{m,n}$ and $q = ((n+N-m))_N$.

Now, for the 3D Euclidean distance $d_{q,v,k}$, first we need to obtain the distance in X-Y plane between one transmitted array-element and one received array-element. Due to the layout of QF-UCA antennas, shifting and rotation of inner-UCA coordinates, differentiated by distinct azimuths under inter-UCA coordinate system, are involved when determining the Euclidean distance.

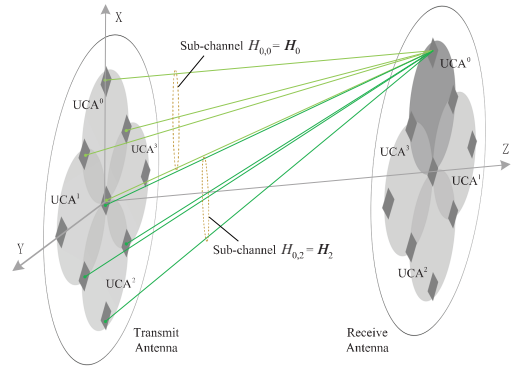

Fig. 2. The equivalent channel gains for transmit signals with 9-element QF-UCA array antennas.

For the rotation issue, we denote by $\varphi_q$ the rotation angle of UCA$^n$ relative to UCA$^m$, written as follows:

$$\varphi_q = \tilde{\Theta}_n - \Theta_m = \frac{2\pi n}{N} - \frac{2\pi m}{M}, \quad (8)$$

where $\tilde{\Theta}_n = 2\pi n/N$ and $\Theta_m = 2\pi m/M$ denote the azimuth of the UCA cells involved at the transmitter and receiver, respectively. The parameter $q = ((n+N-m))_N$ and we have $\varphi_q = 2\pi q/N$ with $M = N$. Mapping onto the coordinate system where UCA$^m$ locates, the azimuth of array-element indexed with $k$ on UCA$^n$ can be derived as $\psi_k + \varphi_q$.

For the shifting issue, we denote by $a_q$ and $b_q$ the distance offsets from the geometric center of UCA$^n$ to that of UCA$^m$ along the y-axis and x-axis, respectively. Combined with the rotation azimuth $\varphi_q$, we have two relative displacements regarding to the center point as follows:

$$\begin{cases} a_q = -R_Q \sin\varphi_q; \\ b_q = -R_Q(1 - \cos\varphi_q). \end{cases} \quad (9)$$

In addition, we denote by $\phi_v$ and $D$ the inner-UCA azimuth of the array-element indexed with $v$ and the vertical distance between transmitter and receiver, respectively. Then, the distance $d_{q,v,k}$ is derived as shown in (10). In (10),

$$h_{q,v,k} = \frac{\beta\lambda}{4\pi D}\exp(\frac{-j2\pi(D+\frac{R_t^2}{2D})}{\lambda})\underbrace{\frac{1}{L_v}\exp\{\frac{-j\pi DB_{q,v}^2}{\lambda R_t^2}\}}_{\mathscr{A}_{q,v}}\exp\{j\overbrace{\frac{2\pi B_{q,v}}{\lambda}}^{\mathscr{B}_{q,v}}\cos(\psi_k-\phi_v+\varphi_q+\alpha_{q,v})\}. \tag{15}$$

$$\begin{aligned}
h_{v,l}^{m,p} &= \frac{\hbar e^{j\Theta_m p}}{\sqrt{KN}}\sum_{q=0}^{N-1}\mathscr{A}_{q,v}\,e^{j\varphi_q p}\sum_{k=0}^{K-1}e^{j\mathscr{B}_{q,v}\cos(\psi_k-\phi_v+\varphi_q+\alpha_{q,v})}e^{j\psi_k l} \\
&= \frac{\sqrt{K}\hbar e^{j\Theta_m p}}{\sqrt{N}}\sum_{q=0}^{N-1}\Bigg[\mathscr{A}_{q,v}\,e^{j\varphi_q p}e^{-j\alpha_{q,v}l}e^{j(\phi_v-\varphi_q)l} \\
&\qquad\qquad\qquad\times \frac{1}{2\pi}\int_{-\pi}^{\pi}e^{j\mathscr{B}_{q,v}\cos(\overbrace{\psi_k-\phi_v+\varphi_q+\alpha_{q,v}}^{\varphi})}e^{j(\psi_k-\phi_v+\varphi_q+\alpha_{q,v})l}\mathrm{d}\varphi\Bigg] \\
&= \frac{(j)^l\sqrt{K}\hbar e^{j\Theta_m p}}{\sqrt{N}}\sum_{q=0}^{N-1}\mathscr{A}_{q,v}\,e^{j(\varphi_q p-\alpha_{q,v}l)}e^{j(\phi_v-\varphi_q)l}J_l(\mathscr{B}_{q,v}) \\
&\approx \frac{(j)^l\sqrt{K}\hbar e^{j\Theta_m p}}{\sqrt{N}}\sum_{q=0}^{N-1}\mathscr{A}_{q,v}\,e^{j(\varphi_q p-\alpha_{q,v}l)}e^{j(\phi_v-\varphi_q)l}J_l\left(\underbrace{\frac{2\pi R_t\sqrt{4R_Q^2\sin^2\frac{\varphi_q}{2}+R_r^2}}{\lambda D}}_{\mathscr{B}_q}\right).
\end{aligned} \tag{16}$$

$$\begin{aligned}
h_{v,q,l}^{m,p} &= \frac{(j)^l\sqrt{K}\,\hbar J_l(\mathscr{B}_q)}{\sqrt{N}}\mathscr{A}_{q,v}\,e^{j\Theta_m p}\,e^{j(\varphi_q p-\alpha_{q,v}l)}e^{j(\phi_v-\varphi_q)l} \\
&= \underbrace{\frac{\beta\lambda\sqrt{K}\,J_l(\mathscr{B}_q)}{4\pi D\sqrt{N}}}_{|h|_q}(j)^l\,e^{\frac{-j2\pi(D+\frac{R_t^2}{2D})}{\lambda}}e^{j\Theta_m p}e^{j\varphi_q p}L_v^{-1}e^{\frac{-j\pi DB_{q,v}^2}{\lambda R_t^2}}\,e^{-j\alpha_{q,v}l}e^{j(\phi_v-\varphi_q)l}.
\end{aligned} \tag{17}$$

$$B_{q,v} = \frac{R_t\sqrt{4R_Q^2\sin^2\frac{\varphi_q}{2}+4R_Q R_r\sin\frac{\varphi_q}{2}\cos(\phi_v-\frac{\varphi_q}{2})+R_r^2}}{D}, \tag{11}$$

and $\alpha_{q,v}$ satisfies the following constraints:

$$\begin{cases} \sin(\alpha_{q,v}) = \frac{2R_Q R_t\sin\frac{\varphi_q}{2}\sin(\phi_v-\frac{\varphi_q}{2})}{DB_{q,v}}; \\ \cos(\alpha_{q,v}) = \frac{2R_Q R_t\sin\frac{\varphi_q}{2}\cos(\phi_v-\frac{\varphi_q}{2})+R_r R_t}{DB_{q,v}}. \end{cases} \tag{12}$$

Therefore, the signal received by the array-element indexed with $v$ on the receive UCA$^m$ cell, denoted by $r_{m,v}$, can be derived as

$$\begin{aligned}
r_{m,v} &= \sum_{n=0}^{N-1}r_{m,v,n} = \sum_{n=0}^{N-1}\sum_{l=1-\lfloor\frac{K}{2}\rfloor}^{\lfloor\frac{K}{2}\rfloor}h_{v,l}^{m,n}\sum_{p=1-\lfloor\frac{N}{2}\rfloor}^{\lfloor\frac{N}{2}\rfloor}e^{j\frac{2\pi n}{N}p}s_{p,l} \\
&= \sum_{l=1-\lfloor\frac{K}{2}\rfloor}^{\lfloor\frac{K}{2}\rfloor}\sum_{p=1-\lfloor\frac{N}{2}\rfloor}^{\lfloor\frac{N}{2}\rfloor}h_{v,l}^{m,p}s_{p,l},
\end{aligned} \tag{13}$$

where

$$h_{v,l}^{m,p} = \frac{e^{j\Theta_m p}}{\sqrt{KN}}\sum_{q=0}^{N-1}e^{j\varphi_q p}\sum_{k=0}^{K-1}e^{j\psi_k l}h_{q,v,k}. \tag{14}$$

The parameter $h_{q,v,k}$ is obtained by substituting (10) into (7), as shown in (15). We now conduct an analyse of $h_{v,l}^{m,p}$ rewritten as (16), where the term $J_l(\cdot)$ represents the $l$-order Bessel function in integral form. Furthermore, the gain regarding the transmitted array-element indexed with $k$, denoted by $h_{v,q,l}^{m,p}$, is given by (17) in the next page to show the amplitude and phase change of the signals.

Then, we focus on analyzing the change of amplitude and phase based on $h_{v,q,l}^{m,p}$. The change rule in amplitude of the signal transmitted by array-element indexed with $k$ is as follows: regardless of the $l$-order Bessel term, $|h|_{q,v}$ changes as a result of the constant term $\beta\lambda\sqrt{K}/(4\pi D\sqrt{N})$ and the $l$-order Bessel function depends on the inner-UCA azimuth $\phi_v$ as well as inter-UCA azimuth $\varphi_q$. The change rule in phase is as follows: regardless of the constant term, the phase part depends on the values of inter-UCA OAM mode $p$, inner-UCA OAM mode $l$, inter-UCA azimuth $\tilde{\Theta}_n$ and $\Theta_m$, and inner-UCA azimuth $\phi_v$ and $\psi_k$.

Furthermore, we show a specific case corresponding to the sub-channel between a pair of UCA cells with the same size and index as follows:

**A Specific Case:** Consider a specific case when $R_r = R_t$ and $q = 0$. Substituting the above conditions into (8), (9), (10), (11), and (15), we have

$$\begin{cases} \varphi_q = 0,\ a_0 = 0,\ b_0 = 0; \\ B_{q,v} = \frac{R_t R_r}{D^2},\ \alpha_{q,v} = \frac{\pi}{2}-\phi_v; \\ d_{0,v,k} = \sqrt{D^2+R_r^2+R_t^2-2R_r R_t\cos(\phi_v-\psi_k)}; \\ h_{0,v,k} = \frac{\beta\lambda}{4\pi DL_v}e^{\frac{-j2\pi D}{\lambda}}e^{\frac{-j\pi(R_r^2+R_t^2)}{\lambda D^2}}e^{\frac{j2\pi R_r R_t}{\lambda D^2}\cos(\phi_v-\psi_k)}, \end{cases} \tag{18}$$

where the amplitude of $h_{0,v,k}$ represented by $|h_{0,v,k}|$ contains an azimuth-free term of $\beta\lambda/(4\pi D)$ and a term of sharing frequency $L_v$. The cosine component in phase part only depends on the azimuth difference between $\phi_v$ and $\psi_k$. Thus, based on the periodicity of cosine functions, the rest matrix factor, denoted by $\boldsymbol{H}_{cir}$, is a form of circulant matrix as

$$\boldsymbol{H}_0 = \boldsymbol{L}^{-1} \times \boldsymbol{H}_{cir}$$
$$= \boldsymbol{L}^{-1} \times \begin{bmatrix} h_0 & h_1 & \cdots & h_{K-1} \\ h_{K-1} & h_0 & \cdots & h_{K-2} \\ \vdots & \vdots & \ddots & \vdots \\ h_{K-V+1} & h_{K-V+2} & \cdots & h_{K-V} \end{bmatrix}. \quad (19)$$

Moreover, items $h_{k_1}$ and $h_{k_2}$ in the first row of $\boldsymbol{H}_{cir}$ are the same when $k_1+k_2=K$ due to the axisymmetry. In the case of $V=K$, $\boldsymbol{H}_{cir}$ is a symmetric circulant matrix. As for the sub-channel $\boldsymbol{H}_0 \in \mathbb{C}^{(V \times K)}$, its corresponding transmit and receive UCA cells are parallel and aligned to each other as well.

For the symmetric circulant matrix, there is a good property in connection with unitary decomposition that the unitary similarity transformation for $\boldsymbol{H}_{cir}$ produces a diagonal matrix. Specifically, using IDFT matrix $\boldsymbol{W}$ and DFT matrix $\boldsymbol{W}^H$ we have

$$\boldsymbol{W}^H \times \boldsymbol{H}_{cir} \times \boldsymbol{W} = \mathrm{diag}(\boldsymbol{h}^F), \quad (20)$$

where the new vector $\boldsymbol{h}^F \in \mathbb{C}^{(N \times 1)}$ is equal to the DFT of $\boldsymbol{h} = [h_0, h_{K-1}, \ldots, h_2, h_1]^T$ as

$$\boldsymbol{h}^F = \sqrt{K}\boldsymbol{W}^H \times \boldsymbol{h}. \quad (21)$$

### C. Two-dimension DFT Based OAM Demodulation (TOD)

After passing through the splitter, received signals output from $N_r$ array-elements turn into $NV$ signals, which are divided into $N$ branches with each of them containing $V$ items. For different transmit UCA cells, the corresponding sub-channels arranged in matrix $\boldsymbol{H}$ show us a characteristic of circular shifts. Accordingly, as shown in Fig. 1, useful signals received by those shared array-elements are identical after linear superposition, and thus they can be multi-sected directly in accordance with the co-use number of array-elements.

Based on (16) and (17), we can rewrite $r_{m,v}$ in (13) as follows:

$$r_{m,v} = \frac{\sqrt{K}\hbar}{\sqrt{N}} \sum_{p=1-\lfloor \frac{N}{2} \rfloor}^{\lfloor \frac{N}{2} \rfloor} e^{j\Theta_m p} \sum_{l=1-\lfloor \frac{K}{2} \rfloor}^{\lfloor \frac{K}{2} \rfloor} \left[ (j)^l \right.$$
$$\left. \times \sum_{q=0}^{N-1} \mathscr{A}_{q,v} J_l(\mathscr{B}_q) e^{j(\varphi_q p - \alpha_{q,v} l)} e^{j(\phi_v - \varphi_q)l} s_{p,l} \right]. \quad (22)$$

Then, taking $m$ of UCA$^m$ cell as the subscript index, the receive signal vector, denoted by $\boldsymbol{R}$, is derived in (23), where $\boldsymbol{r}_m = [r_{m,1}, \cdots, r_{m,V}]^T$ is the signal vector corresponding to UCA$^m$ cell.

Then, consider the property of $\boldsymbol{W}_N = [\boldsymbol{w}_0, \boldsymbol{w}_1, \ldots, \boldsymbol{w}_{N-1}]^T$ that

$$\boldsymbol{w}_{n_1} \cdot \boldsymbol{w}_{n_2} = \boldsymbol{w}_{((n_1+n_2))_N}/\sqrt{N}, \quad (24)$$

and the equivalence regarding matrix operations from the intercommunity between circulant matrices and cyclic shifts in circular convolution [47] that

$$\boldsymbol{C} \times \boldsymbol{W}_N = \sqrt{N} \, \boldsymbol{W}_N \times \mathrm{diag}(\boldsymbol{c} \times \boldsymbol{W}_N), \quad (25)$$

where $\boldsymbol{C}$ represents a circulant matrix and $\boldsymbol{c}$ is the first row of $\boldsymbol{C}$, we can derive the signal vector $\boldsymbol{R}$ corresponding to inter-UCA cells through the LOS channels as follows:

$$\boldsymbol{R} = \frac{1}{\sqrt{N}} \begin{bmatrix} \boldsymbol{E} & \boldsymbol{E} & \cdots & \boldsymbol{E} \\ \boldsymbol{E} & e^{j\frac{2\pi}{N}}\boldsymbol{E} & \cdots & e^{j\frac{2\pi(N-1)}{N}}\boldsymbol{E} \\ \vdots & \vdots & \ddots & \vdots \\ \boldsymbol{E} & e^{j\frac{2\pi(N-1)}{N}}\boldsymbol{E} & \cdots & e^{j\frac{2\pi(N-1)^2}{N}}\boldsymbol{E} \end{bmatrix}$$
$$\begin{bmatrix} \mathscr{H}_0 & & & \\ & \ddots & & \\ & & \mathscr{H}_p & \\ & & & \ddots \\ & & & & \mathscr{H}_{N-1} \end{bmatrix} \begin{bmatrix} \boldsymbol{W}\boldsymbol{s}_0 \\ \vdots \\ \boldsymbol{W}\boldsymbol{s}_p \\ \vdots \\ \boldsymbol{W}\boldsymbol{s}_{N-1} \end{bmatrix}, \quad (26)$$

where $\mathscr{H}_p$ is given as follows:

$$\mathscr{H}_p = \begin{bmatrix} \boldsymbol{H}_0 \\ \boldsymbol{H}_1 \\ \vdots \\ \boldsymbol{H}_q \\ \vdots \\ \boldsymbol{H}_{N-1} \end{bmatrix}^T \underbrace{\begin{bmatrix} \boldsymbol{E} \\ e^{j\frac{2\pi q}{N}}\boldsymbol{E} \\ \vdots \\ e^{j\frac{2\pi pq}{N}}\boldsymbol{E} \\ \vdots \\ e^{j\frac{2\pi(N-1)q}{N}}\boldsymbol{E} \end{bmatrix}}_{\sqrt{N}\boldsymbol{w}_p}. \quad (27)$$

The parameter $\mathscr{H}_p \in \mathbb{C}^{(V \times K)}$ is a linear superposition of $N$ sub-channels ($\boldsymbol{H}_0, \boldsymbol{H}_1$ and so on) and the superposition coefficient vector is the $p$th column, denoted by $\boldsymbol{w}_p$, of $\boldsymbol{W}_N$.

Then, based on the traditional OAM demodulation scheme, we first conduct the phase compensation for the split signal $r_{m,v}$ as follows:

$$\widetilde{x}_{p,v} = \frac{1}{\sqrt{N}} \sum_{m=0}^{N-1} r_{m,v} e^{-j\Theta_m p}, \quad (28)$$

$$\boldsymbol{R} = \boldsymbol{T}_r \boldsymbol{H}^F \boldsymbol{X}$$

$$= \begin{bmatrix} \boldsymbol{r}_0 \\ \boldsymbol{r}_1 \\ \vdots \\ \boldsymbol{r}_{M-1} \end{bmatrix} = \begin{bmatrix} \boldsymbol{H}_0 & \cdots \boldsymbol{H}_q \cdots & \boldsymbol{H}_{N-1} \\ \boldsymbol{H}_{N-1} \boldsymbol{H}_0 & \cdots & \boldsymbol{H}_{N-2} \\ \vdots & \vdots & \ddots & \vdots \\ \boldsymbol{H}_1 & \boldsymbol{H}_2 & \cdots & \boldsymbol{H}_0 \end{bmatrix} \begin{bmatrix} \boldsymbol{x}_0 \\ \boldsymbol{x}_1 \\ \vdots \\ \boldsymbol{x}_{N-1} \end{bmatrix} = \frac{1}{\sqrt{N}} \begin{bmatrix} \boldsymbol{H}_0 & \cdots \boldsymbol{H}_q \cdots & \boldsymbol{H}_{N-1} \\ \boldsymbol{H}_{N-1} \boldsymbol{H}_0 & \cdots & \boldsymbol{H}_{N-2} \\ \vdots & \vdots & \ddots & \vdots \\ \boldsymbol{H}_1 & \boldsymbol{H}_2 & \cdots & \boldsymbol{H}_0 \end{bmatrix} \begin{bmatrix} \boldsymbol{E} & \boldsymbol{E} & \cdots & \boldsymbol{E} \\ \boldsymbol{E} & e^{j\frac{2\pi}{N}}\boldsymbol{E} & \cdots & e^{j\frac{2\pi(N-1)}{N}}\boldsymbol{E} \\ \vdots & \vdots & \ddots & \vdots \\ \boldsymbol{E} & e^{j\frac{2\pi(N-1)}{N}}\boldsymbol{E} \cdots & e^{j\frac{2\pi(N-1)^2}{N}}\boldsymbol{E} \end{bmatrix} \begin{bmatrix} \boldsymbol{W}\boldsymbol{s}_0 \\ \boldsymbol{W}\boldsymbol{s}_1 \\ \vdots \\ \boldsymbol{W}\boldsymbol{s}_{N-1} \end{bmatrix}.$$
$$(23)$$

$$H_{l_0,l}^{p,q} = \frac{1}{\sqrt{V}}|h|_q\,(j)^l\,e^{\frac{-j2\pi(D+\frac{R_t^2}{2D})}{\lambda}}e^{j\varphi_q p}L_v L_v^{-1}\sum_{v=0}^{V-1}e^{\frac{-j\pi DB_{q,v}^2}{\lambda R_t^2}}e^{-j\alpha_{q,v}l}e^{j(\phi_v-\varphi_q)l}e^{-j\phi_v l_0}$$

$$= \frac{1}{\sqrt{V}}|h|_q\,e^{\frac{-j2\pi(D+\frac{R_t^2}{2D})}{\lambda}}e^{j\varphi_q p}(j)^{l_0}e^{-j\frac{\varphi_q}{2}(l+l_0)}\sum_{v=0}^{V-1}(j)^l(j)^{-l_0}e^{\frac{-j\pi DB_{q,v}^2}{\lambda R_t^2}}e^{-j\alpha_{q,v}l}e^{j(\phi_v-\frac{\varphi_q}{2})(l-l_0)}$$

$$= \sqrt{V}|h|_q\,e^{\frac{-j2\pi(D+\frac{R_t^2}{2D})}{\lambda}}e^{j\varphi_q p}(j)^{l_0}e^{-j\frac{\varphi_q}{2}(l+l_0)}\frac{1}{2\pi}\int_{-\pi}^{\pi}(j)^l(j)^{-l_0}e^{\frac{-j\pi DB_{q,\varphi}^2}{\lambda R_t^2}}e^{-j\alpha_{q,\varphi}l}e^{j(\phi_v-\frac{\varphi_q}{2})(l-l_0)}d\varphi$$

$$= \sqrt{V}|h|_q\,e^{\frac{-j2\pi(D+\frac{R_t^2}{2D})}{\lambda}}e^{\frac{-j\pi(4R_Q^2\sin^2\frac{\varphi_q}{2}+R_r^2)}{\lambda D}}e^{j\varphi_q p}(j)^{l_0}e^{-j\frac{\varphi_q}{2}(l+l_0)}$$
$$\times J_{l-l_0}\left(\frac{4\pi R_Q R_r \sin\frac{\phi_q}{2}}{\lambda D}\right)e^{-j\alpha_{q,0}l} - J_{l-l_0}\left(\frac{4\pi R_Q R_r \sin\frac{\phi_q}{2}}{\lambda D}\right)\underbrace{\int_{-\pi}^{\pi}de^{-j\alpha_{q,\varphi}l}}_{\mathscr{F}_{q,l}}$$

$$\approx \begin{cases} \frac{\beta\lambda\sqrt{KV}}{4\pi D\sqrt{N}}\,e^{\frac{-j2\pi(D+\frac{R_t^2}{2D})}{\lambda}}e^{\frac{-j\pi(4R_Q^2\sin^2\frac{\varphi_q}{2}+R_r^2)}{\lambda D}}e^{j\varphi_q p}(j)^l e^{-j\varphi_q l}J_l(\mathscr{B}_q) \\ \qquad\qquad\qquad\qquad \times J_0\left(\frac{4\pi R_Q R_r \sin\frac{\phi_q}{2}}{\lambda D}\right)\left(e^{-j\alpha_{q,0}l} - \mathscr{F}_{q,l}\right), & l=l_0; \\ 0, & l\neq l_0. \end{cases} \qquad (31)$$

---

where $\widetilde{x}_{p,v}$ corresponds to the item of signal vector $\widetilde{\boldsymbol{x}}_p = [\widetilde{x}_{p,0},\cdots,\widetilde{x}_{p,v},\cdots,\widetilde{x}_{p,V-1}]^T$. For all receive UCA cells, such phase compensation yields a separation of signals with inter-UCA OAM modulation. According to (28), the signal vector, denoted by $\widetilde{\boldsymbol{X}}$, after the separation of signals with inter-UCA OAM modulation is derived as follows:

$$\widetilde{\boldsymbol{X}} = \left[\widetilde{\boldsymbol{x}}_0,\cdots,\widetilde{\boldsymbol{x}}_p,\cdots,\widetilde{\boldsymbol{x}}_{N-1}\right]^T = (\boldsymbol{W}_N)^H \times \boldsymbol{R}$$
$$= \left[\mathscr{H}_0 \boldsymbol{W}\boldsymbol{s}_0,\cdots,\mathscr{H}_p \boldsymbol{W}\boldsymbol{s}_1,\cdots,\mathscr{H}_{N-1}\boldsymbol{W}\boldsymbol{s}_{N-1}\right]^T. \quad (29)$$

In this case, $\mathscr{H}_p$ ($0 \leq p \leq N-1$) can be regarded as the equivalent sub-channel matrix for the signals. We name the above-mentioned operation the OAM based inter-UCA demodulation.

Then, for the OAM based inner-UCA demodulation, we first multiply the post-decoding matrix $\boldsymbol{L}$ to the vector $\widetilde{\boldsymbol{x}}_p$ and then the $V$-point DFT is used to decompose the signals. The signal vector after the OAM based inner-UCA demodulation, denoted by $\widetilde{\boldsymbol{s}}_p$, corresponding to the inter-UCA OAM mode $p$ is derived as follows:

$$\widetilde{\boldsymbol{s}}_p = \boldsymbol{W}^H \boldsymbol{L}\widetilde{\boldsymbol{x}}_p = \boldsymbol{W}^H \boldsymbol{L}\mathscr{H}_p \boldsymbol{W}\boldsymbol{s}_p. \quad (30)$$

Based on (17), we approximate the matrix $\boldsymbol{W}^H \boldsymbol{L}\boldsymbol{H}_q \boldsymbol{W}$ in (30) to $\widetilde{\boldsymbol{H}}_{p,q}$ and its elements, denoted by $H_{l_0,l}^{p,q}$ ($0 \leq l_0, l \leq K-1$), of $\widetilde{\boldsymbol{H}}_{p,q}$ are given in (31). We have that the matrix $\widetilde{\boldsymbol{H}}_{p,q}$ is diagonal. The gap, denoted by $\varepsilon$, between $\boldsymbol{W}^H \boldsymbol{L}\boldsymbol{H}_q \boldsymbol{W}$ and $\widetilde{\boldsymbol{H}}_{p,q}$ is derived as follows:

$$\varepsilon = \frac{|\boldsymbol{W}^H \boldsymbol{L}\boldsymbol{H}_q \boldsymbol{W} - \widetilde{\boldsymbol{H}}_{p,q}|^2}{|\boldsymbol{W}^H \boldsymbol{L}\boldsymbol{H}_q \boldsymbol{W}|^2}. \quad (32)$$

Figure 3 shows the gap $\varepsilon$ versus the distance $D$ and the number of array-elements $K$. We can observe that as $D$ increases, $\varepsilon$ decreases. This is because the non-diagonal elements of $\boldsymbol{W}^H \boldsymbol{L}\boldsymbol{H}_q \boldsymbol{W}$ converges 0 when $D$ increases. The gap $\varepsilon$ decreases as the number of array-elements $K$ increases.

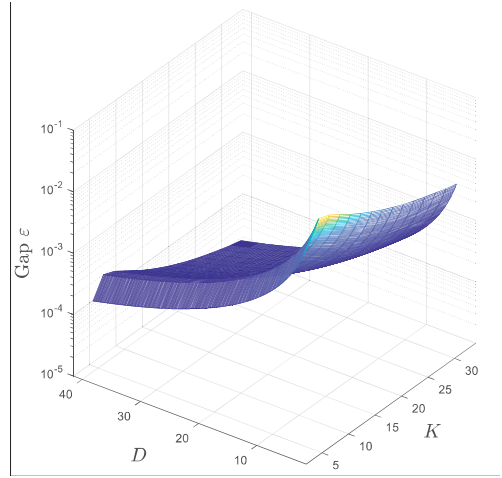

Fig. 3. The gap $\varepsilon$ between $\boldsymbol{W}^H \boldsymbol{L}\boldsymbol{H}_q \boldsymbol{W}$ and $\widetilde{\boldsymbol{H}}_{p,q}$ versus $D$ and $K$.

This is because that the approximations about the Bessel function in (16) and (31) become more accurate as the number of array-elements $K$ increases. More importantly, the gap $\varepsilon$ is small and we can consider that the approximations between $\boldsymbol{W}^H \boldsymbol{L}\boldsymbol{H}_q \boldsymbol{W}$ and $\widetilde{\boldsymbol{H}}_{p,q}$ is effective.

We then adopt $K$-branch maximum-likelihood (ML) detection, which is the mode wise ML detection, for symbol recovery. Traversing all the receive UCA cells along with all OAM modes, we have the signal vector, denoted by $\widehat{\boldsymbol{s}}_p$, corresponding to the $p$th OAM modes for the inter UCA cells as follows:

$$\widehat{\boldsymbol{s}}_p = \arg\min_{\boldsymbol{s}\in\Omega^V}\|\widetilde{\boldsymbol{s}}_p - \boldsymbol{\Lambda}_p \boldsymbol{s}_p\|^2;$$
$$= \arg\min_{s\in\Omega}\sum_{l=1-\lfloor\frac{V}{2}\rfloor}^{\lfloor\frac{V}{2}\rfloor}|\widetilde{\boldsymbol{s}}_p(l) - \Lambda_{p,l}\boldsymbol{s}_p(l)|^2;$$
$$= \arg\min_{s\in\Omega}\sum_{l=1-\lfloor\frac{V}{2}\rfloor}^{\lfloor\frac{V}{2}\rfloor}|\widetilde{\boldsymbol{s}}_p(l) - \Lambda_{p,l}\boldsymbol{s}_p(l)|; \quad (33)$$
$$= [\arg\min_{s\in\Omega}|\widetilde{\boldsymbol{s}}_p(l) - \Lambda_{p,l}\boldsymbol{s}_p(l)|,\cdots,$$
$$\arg\min_{s\in\Omega}|\widetilde{\boldsymbol{s}}_p(l) - \Lambda_{p,l}\boldsymbol{s}_p(l)|],$$

where $\boldsymbol{\Lambda}_p$ is a diagonal matrix corresponding to $p$ ($1-\lfloor\frac{M}{2}\rfloor \leq$

$p \leq \lfloor \frac{M}{2} \rfloor$) and its diagonal elements, denoted by $\Lambda_{p,q,l}$, are derived as follows:

$$\Lambda_{p,l} = \frac{1}{\sqrt{N}} \sum_{q=0}^{N-1} e^{j\frac{2\pi}{N}pq} H_{l,l}^{p,q}. \quad (34)$$

Using our proposed TOD scheme, the signals for different OAM modes are recovered with low complexity.

Then, we investigate the spectrum efficiency of our radio vortex communications. When transmitted and received antennas as well as all their sub-UCAs with same indices are aligned with each other, the spectrum efficiency can be derived as follows:

$$\sum_{p=1-\lfloor \frac{M}{2} \rfloor}^{\lfloor \frac{M}{2} \rfloor} \sum_{l=1-\lfloor \frac{V}{2} \rfloor}^{\lfloor \frac{V}{2} \rfloor} \log_2\left(1 + \frac{|\Lambda_{p,l}|^2 |s_p(l)|^2}{\sigma_{p,l}^2}\right), \quad (35)$$

where $\sigma_{p,l}^2$ is the noise variance corresponding to the $p$th OAM mode of inter-UCA cells and the $l$th OAM mode of inner-UCA cells.

For the single-loop UCA based multiplexing transmission, here the spectrum efficiency is also given as follows:

$$\sum_{l=1-\lfloor N_r/2 \rfloor}^{\lfloor N_r/2 \rfloor} \log_2\left(1 + \frac{|\Lambda_{0,l}|^2 |s_0(l)|^2}{\sigma_l^2}\right), \quad (36)$$

where $\sigma_l^2$ is the noise variance of $l$th OAM mode for the single-loop UCA based OAM multiplexing. To achieve the same spectrum efficiency, there is smaller number of antenna elements using our proposed QF-UCA system as compared with the OAM and MIMO systems. Also, the advantage of using our proposed system is that the OAM modes in our proposed OAM multiplexing based QF-UCA can be recovered without interference and there exists interference among different OAM modes for the OAM and MIMO system. The decomposition complexity of OAM multiplexing based QF-UCA is lower than that of OAM and MIMO system using ML detection.

## IV. PERFORMANCE EVALUATION

In this section, we proceed with numerical evaluation to analyze and evaluate the system performance of our proposed radio vortex multiplexing transmission. First, we compare spectrum efficiencies between the two-dimension OAM multiplexing scheme and the single-loop UCA based OAM scheme. The corresponding spectrum efficiency gains are also given. Regarding the antenna parameters, we consider the QF-UCA antenna with 4-element and 8-element UCA cells. The constant $\beta$ is set to 1 [42]. The distance between the transmitter and receiver is 100m. The radii of transmit UCA cell, receive UCA cell, and QF-UCA antennas are the same. Meanwhile, the relationship between spectrum efficiency gain and antenna layout in four radii settings is also discussed. Throughout the evaluation, we set the communication system operating at the 5.8GHz frequency band and assume transceivers are aligned with each other. The practical systems for QF-UCA antenna will be implemented in the future.

Figure 4 depicts the spectrum efficiencies of OAM multiplexing transmission based on the QF-UCA and the single-loop UCA antennas for 9 elements in four radii settings, respectively. The spectrum efficiency of single-input-single-output (SISO) wireless communication is obtained using the LOS channel model. The power is averagely allocated to OAM modes for OAM multiplexing transmission. As illustrated in Fig. 4, without additional power and frequency consumption, the spectrum efficiency of the 9 array-elements QF-UCA antenna based OAM multiplexing transmission (the legend with "9-element QF-UCA") is higher than those of the single-loop UCA with 9 array-elements (the legend with "9-element UCA") and traditional single antenna by a magnitude of 9 (the legend with "9-SISO"). The spectrum efficiency of 9-SISO is equal to the Shannon limit. This is because the QF-UCA antenna based OAM multiplexing transmission can provide more orthogonal OAM modes, i.e., the number of OAM modes is larger than the number of array elements. The results show a new way to achieve more orthogonal multiplexings beyond the multiplexings of traditional multiple antennas wireless communication systems in correlated channel scenarios. Furthermore, the obtained spectrum efficiency of 9 array-elements QF-UCA antenna based OAM multiplexing transmission is even larger than that of single-loop UCA with 16 array-elements (the legend with "16-element UCA"). This is because more low-order OAM modes can be jointly used in QF-UCA antenna based OAM multiplexing transmission while the OAM beams of high order OAM modes diverge and thus resulting in relatively low spectrum efficiency. When the power is averagely allocated to OAM modes for OAM multiplexing transmission, the spectrum efficiency of our proposed QF-UCA based OAM multiplexing transmission is first larger than and then smaller than that of single-loop UCA based OAM multiplexing transmission, as SNR increases. This is because there are more effective OAM modes for the single-loop 16-element UCA based OAM multiplexing transmission as compared with that for the 9-element QF-UCA based OAM multiplexing transmission when the radius becomes large.

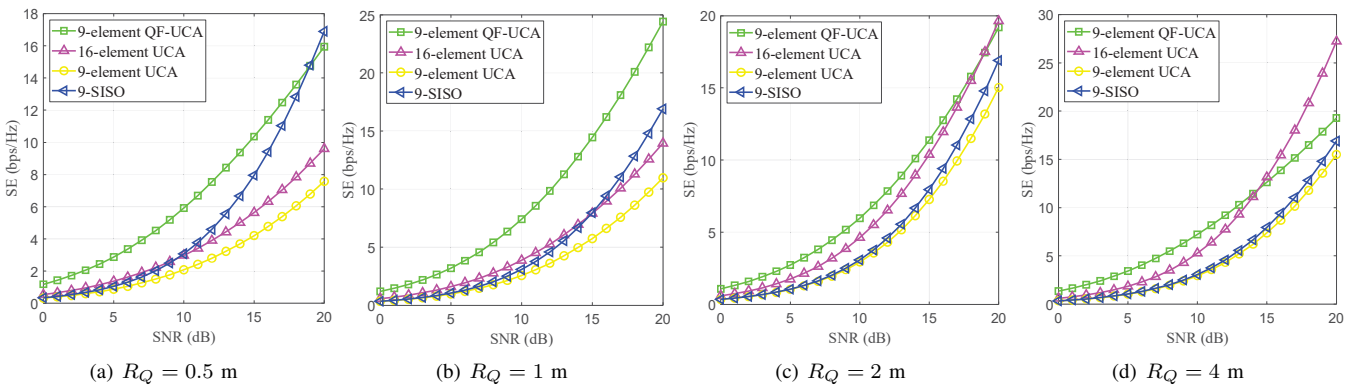

(a) $R_Q = 0.5$ m  (b) $R_Q = 1$ m  (c) $R_Q = 2$ m  (d) $R_Q = 4$ m

Fig. 4. Spectrum efficiencies of the radio vortex multiplexing transmission based on the QF-UCA and the single-loop UCA antennas for 9 elements (with $R_Q = 0.1$m, 0.5m, 1m, and 4m, respectively).

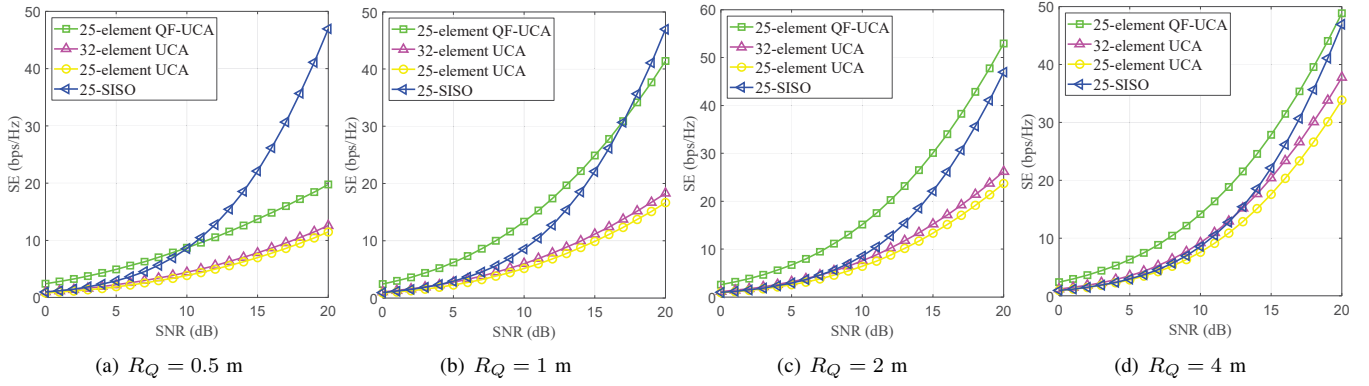

Fig. 5. Spectrum efficiencies of the radio vortex multiplexing transmission based on the QF-UCA and the single-loop UCA antennas for 25 elements (with $R_Q = 0.1$m, 0.5m, 1m, and 4m, respectively).

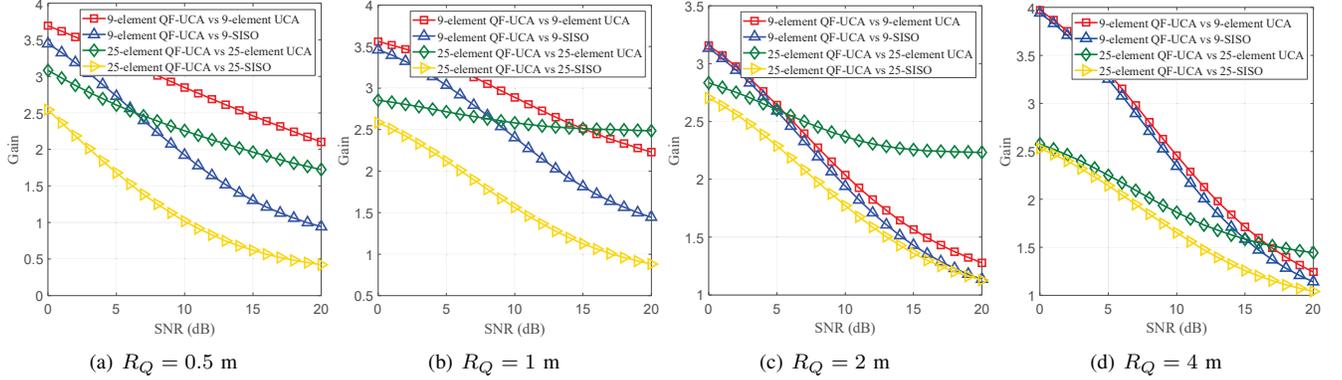

Fig. 6. Spectrum efficiency gains of performing our proposed array-elements QF-UCA antenna based OAM multiplexing transmission versus traditional single-dimension schemes.

Figure 5 shows the spectrum efficiencies of OAM multiplexing transmission based on the QF-UCA and the single-loop UCA antennas for 25 elements in four radii settings, respectively. As illustrated in Fig. 5, without additional power and frequency consumption, compared with the single-loop UCA with 25 array-elements (the legend with "25-element UCA"), the 25 array-elements QF-UCA antenna based OAM multiplexing transmission (the legend with "25-element QF-UCA") can also achieve much higher spectrum efficiency. The obtained SE of single-loop UCA with 32 array-elements (the legend with "32-element UCA") is approximately same as that of single-loop UCA with 25 array-elements (the legend with "25-element UCA"). This is because the large divergence of high-order OAM beams results in low spectrum efficiency. Also, the 25 array-elements QF-UCA antenna based OAM multiplexing transmission possesses a superior SE to that of traditional single antenna by a magnitude of 25 (the legend with "25-SISO") when SNR is small with large values of radii. This is also because the QF-UCA antenna based OAM multiplexing transmission can provide more orthogonal OAM modes with its number exceeding the array-element number.

Also, the spectrum efficiency of 25-element QF-UCA based OAM multiplexing transmission is smaller than that of 25-SISO when SNR is large at small radii. This is because the number of effective OAM modes, which can be used for transmission, is small when the radius is small. The results also confirm the way to investigate more orthogonal multiplexings beyond that of traditional multiple antennas wireless communication systems in correlated channel scenarios. The relationship between the spectrum efficiency of OAM multiplexing transmission and the number of array-elements for the QF-UCA antenna structure, which is related to $R_Q$, is another important question and will be investigated in the future.

Figure 6 shows the spectrum efficiency gains of performing our proposed array-elements QF-UCA antenna based OAM multiplexing transmission versus traditional single-dimension schemes. For relatively less number of array-elements such as 9-element cases, the SE gains of our proposed array-elements QF-UCA antenna based OAM multiplexing transmission over single-loop 9-element UCA and 9 times traditional single antenna are almost larger than 1 for $R_Q = 0.5$m, 1m, 2m, and 4m. For relatively large number of array-elements such as 25-element cases, the SE gains of our proposed array-elements QF-UCA antenna based OAM multiplexing transmission over single-loop 25-element UCA and 25 times traditional single antenna are larger than 1 for $R_Q = 1$m, 2m, and 4m. For $R_Q = 0.5$m, the traditional multiple antennas based scheme can achieve larger SE than that of our proposed array-elements QF-UCA antenna based OAM multiplexing transmission in relatively high SNR region. Thus, further works can aim to exploit the optimal design in terms of number of array-elements for such quasi-fractal structures based array antenna, along with the multi-dimension OAM mode multiplexing scheme, to maximize the spectrum efficiency of radio vortex communications. It is an important question to derive the optimal $R_Q$ to obtain the maximum spectrum efficiency and we would like to leave the full study of this problem in the future research.

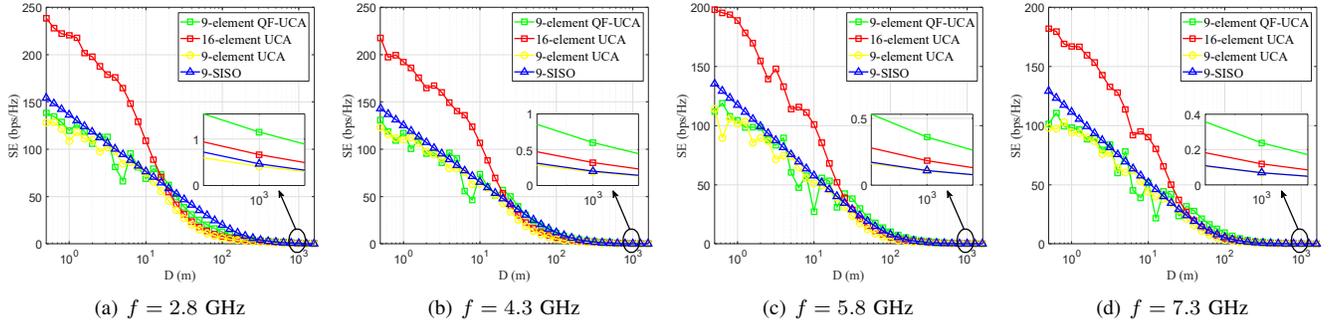

Fig. 7. Spectrum efficiencies of the radio vortex multiplexing transmission based on the QF-UCA and the single-loop UCA antennas for 9 elements (with $f = 2.8$ GHz, $4.3$ GHz, $5.8$ GHz, and $7.3$ GHz, respectively).

Figure 7 shows the spectrum efficiencies versus distance of OAM multiplexing transmission based on the QF-UCA and the single-loop UCA antennas. In Fig. 7, the radius of QF-UCA is 0.5m and the SNR is 15dB. As illustrated in Fig. 7, the spectrum efficiencies of the single-loop UCA with 9 array-elements, the single-loop UCA with 16 array-elements, traditional single antenna by a magnitude of 9, and the 9 array-elements QF-UCA antenna based OAM multiplexing transmission decrease as the distance increases. The obtained spectrum efficiency of 9 array-elements QF-UCA antenna based OAM multiplexing transmission is larger than those of three antenna structures at relatively high frequencies when the power is averagely allocated to OAM modes for OAM multiplexing transmission when the distance is larger than 25m. This is because more effective orthogonal streams are used in QF-UCA antenna based OAM multiplexing transmission thus resulting in relatively high spectrum efficiency, as compared with those of other antenna structures. With the distance increasing, the obtained spectrum efficiency of 9 array-elements QF-UCA antenna based OAM multiplexing transmission is larger than those of three antenna structures for OAM multiplexing transmission. When the distance is small, the obtained spectrum efficiency of 9 array-elements QF-UCA antenna based OAM multiplexing transmission may be smaller than those of three antenna structures at relatively high frequencies. This is because the distance $D$ has a great impact on the transmission of OAM modes for the QF-UCA based OAM multiplexing transmission when the frequency is large.

## V. CONCLUSIONS

In this paper, we proposed the QF-UCA antennas based two-dimension OAM multiplexing to achieve orthogonal OAM modes, the number of which is larger than the number of array-elements, which is beyond the traditional concept of multiple antennas based wireless communications. In particular, we designed the QF-UCA geometric antenna structure by orderly deploying the array-elements in a quasi-fractal layout. Based on the quasi-fractal dimension, we developed the two-dimension DFT based OAM modulation and demodulation schemes to transmit multiple OAM signals. Numerical results validated our proposed QF-UCA antennas based OAM transmission, showing that the obtained spectrum efficiency is generally larger than that of the single-loop UCA based radio vortex transmission. This paper makes a bridge to efficient OAM based wireless communications by considering the multiplexing scheme combined with the antenna geometry, thus increasing the available number of orthogonal OAM modes beyond the number of array-elements which opens the new concept of mode-number over element-number based highly efficient orthogonal transmission.

## APPENDIX A

For the design of QF-UCA antenna, the focus is about the shared array-elements for UCA cells. Considering that array-elements are uniformly distributed in UCA cell, there are some requirements on the layout of QF-UCA antennas, including the range of $R$, $V$, and the number of shared array-elements.

**Case 1:** Since two non-overlapping circles are with at most two points of intersection, we first consider a critical case where the number of shared array-elements for two adjacent UCA cells is 1. Then, we can derive that $\frac{R}{R_Q} = \sin\frac{\pi}{N}$ and $V$ meets

$$V = i\frac{2N}{N-2}, \tag{37}$$

where $i$ and $K$ are both positive integers. Also, we denote by $V_{min}$ the minimum value of $V$. Thus, the diagonal elements of sharing frequency matrix $\boldsymbol{L}$, denoted by $\text{diag}(\boldsymbol{L})$, can be obtained by the circulant shift of the following vector:

$$[\,2,\, \overbrace{1,\cdots 1}^{i-1\ \text{numbers}},\, 2,\, \underbrace{1,\cdots 1}_{N_g\ \text{numbers}}\,]^T, \tag{38}$$

where $N_g = V - i - 1$.

**Case 2:** Next, we consider the case of $\sin\frac{\pi}{N} < \frac{R}{R_Q} < 1$. Assume that there are no shared array-elements arranged for non-adjacent UCA cells. When the two adjacent UCA cells

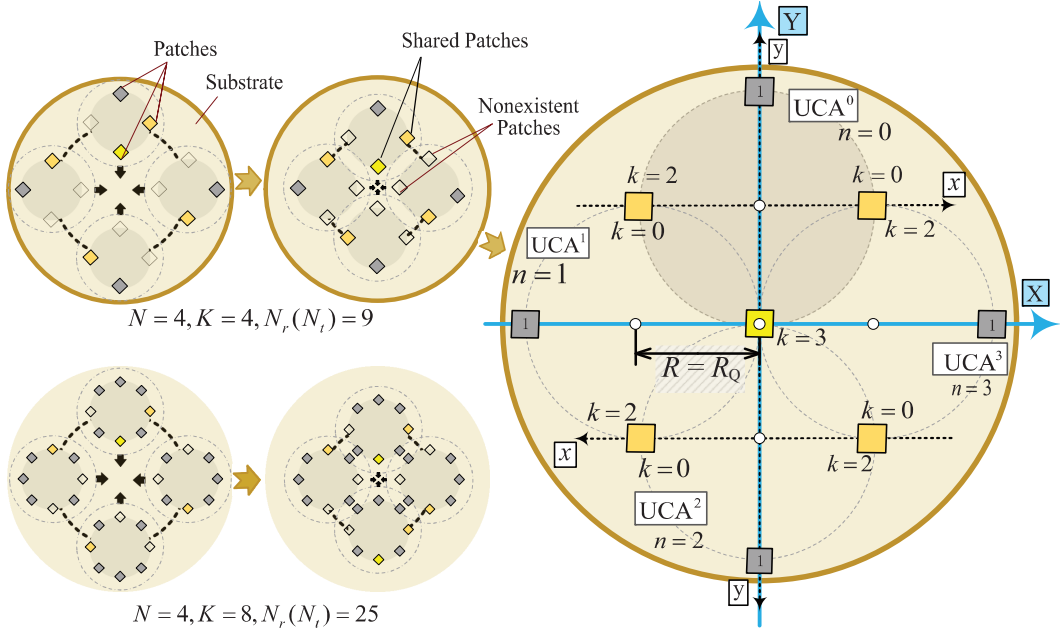

Fig. 8. QF-UCA antennas with 9 and 25 array-elements.

$$[\,N,\; \underbrace{1,\cdots 1}_{\frac{V}{V_{min}}-1 \text{ numbers}},\; 2,\; \underbrace{1,\cdots 1}_{\frac{V}{V_{min}}-1 \text{ numbers}},\; ,\cdots 2,1,\cdots 1,\; \underbrace{1,\cdots 1}_{\frac{V}{V_{min}}-1+N_g \text{ numbers}},\; 2,\; \underbrace{1,\cdots 1}_{\frac{V}{V_{min}}-1 \text{ numbers}},\; ,\cdots 2,1,\cdots 1\,]^T. \quad (41)$$

$\lfloor \frac{V_{min}-1}{2} \rfloor$ numbers of $2,1,\cdots 1$ ... $\lfloor \frac{V_{min}-1}{2} \rfloor$ numbers of $2,1,\cdots 1$

have two shared array-elements, according to the geometric relationship, $V$ needs to meet

$$V = \begin{cases} i_1 \dfrac{\pi}{\frac{\pi}{2} - \arccos(\frac{R_Q}{R}\sin\frac{\pi}{N}) - \frac{\pi}{N}}; \\ i_2 \dfrac{\pi}{\arccos(\frac{R_Q}{R}\sin\frac{\pi}{N})}, \end{cases} \quad (39)$$

where $i_1$, $i_2$, and $V$ are positive integers. Then, $\text{diag}(\boldsymbol{L})$ can be obtained by the circulant shift of the vector as follows:

$$[\,2,\; \underbrace{1,\cdots 1}_{i_1-1 \text{ numbers}},\; 2,\; \underbrace{1,\cdots 1}_{N_g \text{ numbers}},\; 2,\; \underbrace{1,\cdots 1}_{i_1-1 \text{ numbers}},\; 2,\; \underbrace{1,\cdots 1}_{i_2-1 \text{ numbers}}\,]^T, \quad (40)$$

where $N_g = V - 2i_1 - i_2 - 1$. Moreover, when arranging shared array-elements for non-adjacent UCA cells, more conditions need to be satisfied simultaneously for $V$. Then, the number of 2 in frequency matrix $\boldsymbol{L}$ will increase correspondingly.

**Case 3:** Consider the case of $\frac{R}{R_Q} = 1$, where exists one array-element shared for all UCA cells. Therefore, $V = i\frac{2N}{N-2}$ with $i$ and $V$ both being positive integers. In this case, we have a sharing frequency vector as shown in (41), where $N_g = V_{min} - 2\lfloor \frac{V_{min}-1}{2} \rfloor - 1 = \{0, 1\}$.

As shown in Fig. 8, taking $N = 4$ as an example, we have $V_{min} = 4$ and $V = \{4, 8, 12, \cdots\}$. If $V = 4$ and 8, we have $\boldsymbol{L} = \text{diag}([2, 1, 2, 4])$ and $\text{diag}([2, 1, 1, 1, 2, 1, 4, 1])$ as the result of circulant shift for vectors $[4, 2, 1, 2]^T$ and $[4, 1, 2, 1, 1, 1, 2, 1]^T$ derived from (41), respectively.

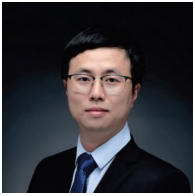

**Wenchi Cheng** received the B.S. and Ph.D. degrees in telecommunication engineering from Xidian University, Xian, China, in 2008 and 2013, respectively. He was a Visiting Scholar with the Department of Electrical and Computer Engineering, Texas A&M University, College Station, TX, USA, from 2010 to 2011. He is currently a Full Professor with Xidian University. He has published more than 200 international journal and conference papers in IEEE JOURNAL ON SELECTED AREAS IN COMMUNICATIONS, IEEE MAGAZINES, IEEE TRANSACTIONS, IEEE INFOCOM, GLOBECOM, and ICC. His current research interests include 6G wireless networks, electromagnetic-based wireless communications, and emergency wireless information. He received the IEEE ComSoc AsiaCPacific Outstanding Young Researcher Award in 2021, the URSI Young Scientist Award in 2019, the Young Elite Scientist Award of CAST, and four IEEE journal/conference best papers. He has served or serving as the Wireless Communications Symposium Co-Chair for IEEE ICC 2022 and IEEE GLOBECOM 2020, the Publicity Chair for IEEE ICC 2019, the Next Generation Networks Symposium Chair for IEEE ICCC 2019, and the Workshop Chair for IEEE ICC 2019/IEEE GLOBECOM 2019/INFOCOM 2020 Workshop on Intelligent Wireless Emergency Communications Networks. He has served or serving as the ComSoc Representative for IEEE Public Safety Technology Initiative, Editor for IEEE Transactions on Wireless Communications, IEEE System Journal, IEEE Communications Letters, and IEEE Wireless Communications Letters.

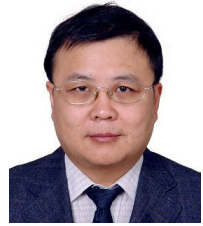

**Hailin Zhang** received his B.S. and M.S. degrees from Northwestern Polytechnic University, Xian, China, in 1985 and 1988, respectively, and his Ph.D. degree from Xidian University in 1991. In 1991, he joined the School of Telecommunications Engineering, Xidian University, where he is a senior professor and the dean of this school. He is also currently the director of the Key Laboratory in Wireless Communications sponsored by the China Ministry of Information Technology, a key member of the State Key Laboratory of Integrated Services Networks, one of the state government specially compensated scientists and engineers, a field leader in telecommunications and information systems at Xidian University, and an associate director of the dean of the National 111 Project. His current research interests include key transmission technologies and standards of broadband wireless communications for 5G and 5G-beyond wireless access systems. He has published more than 150 papers in journals and conferences.

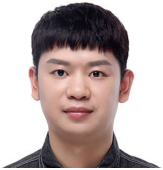

**Haiyue Jing** received the B.S. degree in telecommunication engineering from Xidian University, China, in 2017. He is currently pursuing the Ph.D. degree in telecommunication engineering at Xidian University. His research interests focus on B5G/6G wireless networks, OAM based wireless communications, and LOS MIMO wireless communications.

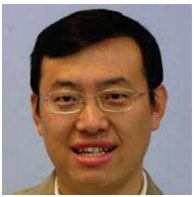

**Wei Zhang** received the Ph.D. degree in electronic engineering from The Chinese University of Hong Kong in 2005. He is currently a Professor with the School of Electrical Engineering and Telecommunications, University of New South Wales, Sydney, NSW, Australia. He has published more than 200 articles and holds five U.S. patents. His research interests include millimetre wave communications and massive MIMO. He is the Vice Director of the IEEE ComSoc Asia Pacific Board. He serves as an Area Editor for the IEEE TRANSACTIONS ON WIRELESS COMMUNICATIONS and the Editor-in-Chief for Journal of Communications and Information Networks.

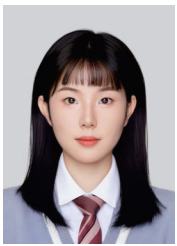

**Keyi Zhang** received the B.S. and M.S. degrees in telecommunication engineering from Xidian University, China, in 2019 and 2022, respectively. Her research interests focus on OAM based wireless communications.